\documentclass[twocolumn,showpacs,superscriptaddress,preprintnumbers,amsmath,amssymb]{revtex4}

\usepackage{graphicx}
\usepackage{dcolumn}
\usepackage{bm}
\usepackage{type1cm}
\newcommand{\phizero}{\!{\stackrel{\,_{(0)}}\phi}\!}
\newcommand{\phione}{\!{\stackrel{\,_{(1)}}\phi}\!}
\newcommand{\phitwo}{\!{\stackrel{\,_{(2)}}\phi}\!}
\newcommand{\eq}{\!\!\!=\!\!\!}

\begin{document}

\preprint{KUNS-2071}

\title{Primordial Non-Gaussianity in Multi-Scalar Slow-Roll Inflation}

\author{Shuichiro Yokoyama}
\email{shu@tap.scphys.kyoto-u.ac.jp}
\affiliation{Department of Physics, Kyoto University, Kyoto 606-8502, Japan}

\author{Teruaki Suyama}%
\email{suyama@icrr.u-tokyo.ac.jp}
\affiliation{Institute for Cosmic Ray Research, The University of Tokyo, Kashiwa 277-8582, Japan}

\author{Takahiro Tanaka}
\email{tama@scphys.kyoto-u.ac.jp}
\affiliation{Department of Physics, Kyoto University, Kyoto 606-8502, Japan}
 
\date{\today}

\begin{abstract}
We analyze the non-Gaussianity for primordial curvature perturbations generated in multi-scalar slow-roll inflation model including 
the model with non-separable potential by making use of $\delta N$ formalism.
Many authors have investigated the possibility of large non-Gaussianity for the models with separable potential,
and they have found that the non-linear parameter, $f_{NL}$, is suppressed by the slow-roll parameters.
We show that for the non-separable models 
$f_{NL}$ is given by the product of a factor which is suppressed by the slow-roll parameters 
and a possible enhancement factor 
which is given by exponentials of quantities of $O(1)$. 

\end{abstract}

\maketitle
\section{Introduction}
Inflation has been widely recognized as a standard mechanism for
generating primordial density perturbations which seed 
the structure formation of the universe and 
the cosmic microwave background(CMB) anisotropies.
In the simplest single-field inflationary universe scenario, 
primordial fluctuations are created by
vacuum fluctuations of the inflaton.
However, in constructing realistic inflation models based on
supergravity, 
it seems more natural to consider that 
the energy scale of inflation is much lower 
and that the scalar field may have 
multi-components during inflation. 
The discrimination of the simplest single-field inflation model from the 
other low energy inflation models will be most clearly 
done by the future observation of 
CMB B-mode polarization~\cite{:2006uk,Taylor:2006jw,quiet}. The simplest single-field model
predicts high energy scale of inflation. Thus the amplitude of 
the tensor perturbation is large and is expected to be observed soon. 
In contrast, in the case of low energy inflation models, 
the tensor perturbation is negligibly small. Therefore no primordial 
tensor perturbation will be detected. 

Recently, the non-linearity (non-Gaussianity) of the primordial
perturbations also has been 
a focus of constant attention by many authors
~\cite{Bartolo:2004if,Maldacena:2002vr,Seery:2005wm,Malik:2003mv,Rigopoulos:2003ak,Lyth:2004gb,Langlois:2005ii,Lyth:2005du,Lyth:2005fi,Alabidi:2005qi,Tanaka:2006zp,Komatsu:2001rj}.
The main reason for attracting much attention 
is that meaningful measurement of this quantity, 
which brings us valuable 
information about the dynamics of inflation if detected, 
will become observationally available in near future. 
In order to parameterize the amount of non-Gaussianity of 
primordial perturbations, 
commonly used is a non-linear parameter, $f_{NL}$, 
which is related to the bispectrum of the curvature perturbation~\cite{Komatsu:2001rj}. 

Meanwhile, in the single field slow-roll inflation it is found that 
$f_{NL}$ is suppressed by
slow-roll parameters to undetectable level~\cite{Maldacena:2002vr,Lyth:2005fi}.
But, for example, in the curvaton
scenario~\cite{Moroi:2001ct,Lyth:2001nq}, 
it is predicted by many 
authors~\cite{Lyth:2002my,Bartolo:2003jx,Malik:2006pm,Sasaki:2006kq} 
that there is a possibility of large non-Gaussianity enough to 
be detectable by future experiments, such as PLANCK~\cite{:2006uk}, 
which is expected to detect
the non-linear parameter if 
$\left|f_{NL}\right| \gtrsim 5$~\cite{Komatsu:2001rj}.
In curvaton scenario, primordial curvature perturbations are sourced by 
isocurvature perturbations related to the vacuum fluctuations of a
light scalar field (other than inflaton), called curvaton, 
which is an energetically subdominant component during inflation.
As the energy density of the universe drops after inflation, 
the fraction of this component becomes significant. 
Then, through the process that the curvaton decays into radiation after inflation,
the curvaton isocurvature perturbations are converted 
into curvature (adiabatic) perturbations.
Curvaton scenario predicts nearly scale invariant spectrum as in 
the case of the standard inflation scenario, but a
large value of $f_{NL}$ is possible in this scenario. 

For the multi-scalar field inflation, however, the possibility of
generation of primordial non-Gaussianity 
has been studied only for the models with the separable potential within the slow-roll 
approximation~\cite{Rigopoulos:2005xx,Kim:2006te,Vernizzi:2006ve,Battefeld:2006sz,Battefeld:2007en,Choi:2007su}.
For such models with separable potential, it was predicted that $f_{NL}$
is suppressed by slow-roll parameters as long as slow-roll
conditions are satisfied.

In this paper, we analyze the primordial non-Gaussianity in multi-scalar
field inflation models without specifying the explicit form of 
the potential. What we assume is just the slow-roll conditions. 
To obtain an analytic formula for $f_{NL}$ written in terms of the
potential of scalar fields, 
we apply $\delta N$ formalism
~\cite{Sasaki:1995aw}
 extended to non-linear regime 
\cite{Sasaki:1998ug, Lyth:2005fi}.

In section \ref{correlationfunc} we briefly review the power spectrum
and the bispectrum which is related to the two- and three-point
correlation functions of the curvature perturbations respectively and we
define the non-linear parameter, $f_{NL}$.
In section \ref{deltaNsec} we review how the non-linear parameter can be
described in the $\delta N$ formalism which was proposed in
Ref.~\cite{Lyth:2005fi}. In section
\ref{evaluate} we show how one can obtain an analytic formula for
$f_{NL}$ in terms 
of the potential of the scalar field in slow-roll approximation. 
We also discuss the possibility of generation of a large amplitude of the primordial
non-Gaussianity in multi-scalar slow-roll inflation. 
We give a summary in section \ref{conclusion}.

\section{Non-linear parameter $f_{NL}$ and $\delta N$ formalism} 

\subsection{Power Spectrum and bispectrum}
\label{correlationfunc}
In this subsection, we briefly review the power spectrum and bispectrum of curvature perturbations and 
define the non-linear parameter, $f_{NL}$, following Ref.~\cite{Lyth:2005fi,Lyth:2005du}.
We consider a minimally coupled $D$-component scalar field whose action is given by
\begin{equation}
S_{\rm fields} = - \int d^4x \sqrt{-g} \left[{1 \over 2}g^{\mu\nu}\delta_{IJ}\partial_{\mu}\phi^I
\partial_{\nu}\phi^J + V(\phi)\right]~,
\nonumber
\end{equation}
where $V(\phi)$ represents the potential of the scalar field. 

As a gauge invariant perturbation variable, 
we choose the curvature perturbation on a uniform density hypersurface, $\zeta$. 
If the perturbation is pure Gaussian, its statistical properties are characterized by its power
spectrum, ${\cal P}_{\zeta},$ 
defined by 
\begin{eqnarray}
\langle \zeta_{{\bf k}_1}\zeta_{{\bf k}_2}\rangle \equiv
\delta^{(3)}\left({\bf k}_1+{\bf k}_2\right){2\pi^2 \over k^3_1}{\cal P}_{\zeta}(k_1)~,
\label{power}
\end{eqnarray}
where $\zeta_{{\bf k}}$ represents a Fourier component given by
\begin{equation}
\zeta_{{\bf k}}(t) = {1 \over \left( 2\pi \right)^{3/2}} \int d^3x\, \zeta(t, {\bf x})\exp \left(-i{\bf k}\cdot {\bf x}\right)~.
\nonumber
\end{equation} 
What we consider in this paper is the non-Gaussian curvature perturbation given in the form~\cite{Bartolo:2004if, Maldacena:2002vr},
\begin{eqnarray}
\zeta ({\bf x}) = \zeta_G({\bf x}) -{3 \over 5}f_{NL}\zeta_G^2({\bf x})~,
\label{fNL}
\end{eqnarray}

where $f_{NL}$ is called the non-linear parameter and $\zeta_G$ satisfies Gaussian statistics.
The power spectrum of $\zeta$ is identical to that of $\zeta_G$, i.e. ${\cal P}_{\zeta}={\cal P}_{\zeta_G}$,
and non-linear part, $-{3 \over 5}f_{NL}\zeta_G^2$, affects the higher order correlation functions.
The three-point correlation function is 
characterized by the bispectrum, $B$, defined by 

\begin{equation}
\langle \zeta_{{\bf k}_1}\zeta_{{\bf k}_2}\zeta_{{\bf k}_3} \rangle \equiv
\delta^{(3)}\left({\bf k}_1 + {\bf k}_2 + {\bf k}_3\right)B_{\zeta}(k_1,k_2,k_3)~.
\label{bispectrum}
\end{equation}
In the present case, the bispectrum $B$ is expressed as
\begin{eqnarray*}
B_{\zeta}(k_1,k_2,k_3) &\eq & -{6 \over 5}{f_{NL}\over \left(2\pi\right)^{3/2}}
\Biggl[P_{\zeta}(k_1)P_{\zeta}(k_2) 
\cr &&\qquad\qquad
+ P_{\zeta}(k_2)P_{\zeta}(k_3) 
+ P_{\zeta}(k_3)P_{\zeta}(k_1)\Biggr]~,
\end{eqnarray*}
where $P_{\zeta}(k)=2\pi^2{\cal P}_{\zeta}(k)/k^3$. 

\subsection{$\delta N$ formalism}
\label{deltaNsec}

In this subsection we briefly review the $\delta N$ formalism and show a simple formula for $f_{NL}$,
following Ref.~\cite{Sasaki:1995aw,Sasaki:1998ug} and \cite{Lyth:2005fi}.

The background $e$-folding number between an initial hypersurface at $t=t_*$ and a final hypersurface 
at $t=t_{\rm c}$ is defined by
\begin{equation}
N \equiv \int H dt~.
\nonumber
\end{equation}
Here, we assume that the time derivative of $\phi^I(t)$
 is not independent of $\phi^I(t)$ as in the case of standard slow-roll inflation, and then 
we can regard $N$ as a function of the homogeneous background field configuration 
$\phi^I(t_*)$ on the initial hypersurface at 
$t=t_{*}$ and $\phi^I(t_{\rm c})$ on the final hypersurface at $t=t_{\rm c}$,
\begin{equation*}
N=N(\phi^I(t_{\rm c}),\phi^I(t_*))~.
\end{equation*}

Let us take $t_{*}$ to be a certain time soon after the relevant length 
scale crossed the horizon scale, and 
$t_{\rm c}$ to be a time 
when the complete convergence of background trajectories in the phase space of the $D$-component scalar field has occurred.
At $t>t_{\rm c}$ the history of the universe is labelled by a single parameter.
Then, it is well-known 
that the curvature perturbation on a uniform density hypersurface, $\zeta$,
becomes constant in time on super-horizon scales.
Thus, in the estimation of the spectrum, what we need is only the final value of the curvature perturbation $\zeta(t_{\rm c})$. 
Based on the $\delta N$ formalism, $\zeta$
evaluated at $t=t_{\rm c}$ is given by 
$\delta N(t_{\rm c},\phi^I(t_*))$, 
where we substitute $\phi^I(t_{\rm c})$ with $t_{\rm c}$ 
because we have taken $t_{\rm c}$ to be a time when the background trajectories have converged.
 
The relation between $\zeta$ and $\delta N$ up to the linear order is given in Ref.~\cite{Sasaki:1995aw},
and its non-linear extension is given in Ref.~\cite{Sasaki:1998ug},
(See also \cite{Lyth:2004gb}).   
Up to second order the relation becomes 
\begin{equation}
\zeta(t_{\rm c}) \simeq \delta N(t_{\rm c},\phi^I_{*}) = N^*_{I}\delta
 \phi^I_{*} + {1 \over 2}N^*_{IJ}\delta \phi^I_* \delta \phi^J_*~,
\label{deltaNsecond}
\end{equation}
where $\simeq$ means the use of the super-horizon approximation, 
and $\delta \phi^I_*$ represent the field perturbations on the initial
flat hypersurface at $t=t_*$.
We have also defined $N^*_I=N_I(t_*)$ and $N^*_{IJ}=N_{IJ}(t_*)$ with 
\begin{eqnarray*}
&&N_I(t)\equiv \left.{\partial N(t_{\rm c},\phi^I) 
  \over \partial \phi^I}\right\vert_{\phi^I=\phi^I(t)},\cr
&&N_{IJ}(t) \equiv
 \left.{\partial^2 N(t_{\rm c},\phi^I) 
    \over \partial \phi^I\partial \phi^J}
    \right\vert_{\phi^I=\phi^I(t)}~. 
\end{eqnarray*}

Substituting Eq.~(\ref{deltaNsecond}) into Eqs.~(\ref{power}) 
and (\ref{bispectrum}), we find that the non-linear parameter $f_{NL}$ defined by Eq.~(\ref{fNL}) is given by~\cite{Lyth:2005fi,Vernizzi:2006ve}
\begin{equation}
 - {6 \over 5}f_{NL} \simeq{N_*^{I}N_*^{J}N^*_{IJ} \over 
   \left(N^*_{K}N_*^{K}\right)^2}~,
\label{fNLdeltaN}
\end{equation}
where the indices are lowered and raised by using the Kronecker's delta 
like
\begin{equation}
N^{I} = \delta^{IJ}N_J. 
\nonumber
\end{equation}
Here we have assumed that the field perturbation 
on the initial flat hypersurface, $\delta \phi^I_*$, is Gaussian
\footnotemark[30] \footnotetext[30]{
In Ref.~\cite{Seery:2005wm,Lyth:2001nq}, the authors have calculated the
non-Gaussianity of the field 
perturbations on the initial flat hypersurface, $\delta \phi_*$,
and they concluded that the non-Gaussianity will not be large when the
$"src"$ are satisfied.
Here we assume the slow-roll approximation, so it does not seem so bad 
to neglect the contribution of the non-Gaussianity of 
$\delta \phi_*$ to $f_{NL}$.
},
and we have neglected the logarithmic term appearing in Ref.~\cite{Lyth:2005fi} because this term is suppressed by the power spectrum 
of curvature perturbation, ${\cal P}_{\zeta} \sim 10^{-10}$.

\section{non-linear parameter in multi-scalar slow-roll inflation}
\label{evaluate}
\subsection{Background equations in slow-roll regime}

Using the background $e$-folding number as the time coordinate, 
the background equation is obtained as
\begin{equation}
{d^2 \over dN^2}\phi^I + \left(3 + {1 \over H}{dH \over dN}\right){d\phi^I \over dN} + {V^I \over H^2} = 0~,
\nonumber
\end{equation}
where $V^I = \delta^{IJ}(\partial V / \partial \phi^J)$.
The background Friedmann equation is given by
\begin{eqnarray}
H^2 &\eq& {1 \over 3} \left({1 \over 2}H^2{d \phi^I \over dN}{d \phi_I \over dN} + V(\phi)\right)~.
\nonumber
\end{eqnarray}

We define the slow-roll parameters in terms of the potential of the scalar field as 
\begin{eqnarray}
\epsilon \equiv {1 \over 2}{V^IV_{I} \over V^2}~,\qquad \eta_{IJ} \equiv {V_{IJ} \over V}~.
\nonumber
\end{eqnarray}
Hereinafter, we assume $\epsilon \ll 1$ and $|\eta_{IJ}| \ll 1$, which we 
call "relaxed" slow-roll conditions (RSRC).
\footnotemark[31] \footnotetext[31]
{
In the "standard" slow-roll approximation,
one assume that $\epsilon \approx |\eta_{IJ}| \ll 1$.
Here, we do not assume the relation between the order of $\epsilon$ and that of $\eta_{IJ}$.
}
Under these conditions, the background equations reduce to 
\begin{eqnarray}
{d \phi^I \over dN} &\simeq& -{V^I \over V}~,\label{eom}\\
H^2 &\simeq& {1 \over 3 }V(\phi)~.
\end{eqnarray}

In most cases, the complete convergence of background trajectories in phase
space of the $D$-component scalar field 
occurs after the RSRC are invalidated. 
Under the present approximation, therefore, 
we can not evaluate $\zeta(t_{\rm c})$, the curvature perturbation on 
a uniform density hypersurface after the convergence of trajectories. 
In this paper, we concentrate on the non-Gaussianity of the curvature
perturbation generated during the RSRC phase. 
For this purpose, we introduce $t_{\rm f}$, a time at which the RSRC
 are still satisfied. 
We divide the process of evaluating $\zeta(t_{\rm c})$ into two parts: 
(i) evaluation of $N(t_{\rm c}, \phi^I(t_{\rm f}))$, the 
$e$-folding number to reach $\phizero^{I}(t_{\rm c})$
starting with $\phi^I=\phi^I(t_{\rm f})$, and 
(ii) expressing $\delta\phi^I(t_{\rm f})$ in terms of 
$\delta\phi^I_*$, where we expand the scalar field as
$\phi^I \equiv \phizero^I + \delta \phi^I$. 
As 
we use the $e$-folding number itself as the time coordinate,  
$N+N(N_{\rm c},\phi^I(N))$, by definition, is equal to $N_{\rm c}$ 
and is constant independent of $N$. 
Therefore one can say
\begin{eqnarray}
\zeta(N_{\rm c}) 
\simeq \delta N(N_{\rm c},\phi^I(N_{\rm f})) \approx
N^{\rm f}_I \delta\phi^I_{\rm f}+
{1\over 2}N^{\rm f}_{IJ} \delta\phi^I_{\rm f}\delta\phi^J_{\rm f},
\label{deltaNfend}
\end{eqnarray} 
where 
$\delta\phi^I_{\rm f}$$=$$\delta\phi^I(N_{\rm f})$,
$\delta N(N_{\rm c},\phi^I(N_{\rm f})) =$
$ N(N_{\rm c},\phi^I(N_{\rm f}))$
$ - N(N_{\rm c},\phizero^I(N_{\rm f}))$,
$N^{\rm f}_I=N_I(N_{\rm f})$ and $N^{\rm f}_{IJ}=N_{IJ}(N_{\rm f})$. 
Our formulation does not apply for the curvature perturbation generated 
after the RSRC are violated. 
Postponing the evaluation of this part to the future issue, 
we study how one can express $\delta\phi^I_{\rm f}$ 
in terms of $\delta\phi^I_*$ to the second order perturbation.   
In the succeeding subsection, assuming that 
$N^{\rm f}_I$ and $N^{\rm f}_{IJ}$ are given, we 
calculate the non-linear parameter, $f_{NL}$.

\subsection{Analytic formula for the non-linear parameter}

In order to obtain the non-linear parameter, $f_{NL}$, 
introduced in Eq.~(\ref{fNLdeltaN}), 
we need to evaluate $N^*_I$ and $N^*_{IJ}$. 
Once we obtain the relation between $\delta\phi^I_{\rm f}$
and $\delta\phi^I_{*}$, 
one can express $N^*_I$ and $N^*_{IJ}$ by 
using $N^{\rm f}_I$ and $N^{\rm f}_{IJ}$ 
from the comparison of Eqs.~(\ref{deltaNsecond})
and (\ref{deltaNfend}). 
Thus, we first solve the evolution of $\delta\phi^I(N)$ to the 
second order. 

The scalar field is expanded up to second order as
\begin{equation}
 \phi^I \equiv \phizero^I+
  \delta \phione^I + {1 \over 2} \delta \phitwo^I+ \cdots~. 
\nonumber
\end{equation}
Taking the variation of Eq.~(\ref{eom}), we obtain 
\begin{eqnarray}
{d \over dN}\delta \phione^I(N) &\eq& 
 \delta\phione^J(N) P^I_{~J}(N)~,
\label{deltaphi1}\\
{d \over dN}\delta\phitwo^I(N) &\eq& 
 \delta \phitwo^J(N) P^I_{~J}(N)
\cr &&\quad  
 + \delta\phione^J(N) \delta\phione^K(N) Q^I_{~JK}(N)~,
\label{deltaphi2}
\end{eqnarray}
with
\begin{eqnarray*}
P^I_{~J}(N) & \equiv & \left[-{V^I_{~J} \over V} + {V^I V_J \over V^2}
   \right]_{\phi=\phizero(N)},\nonumber \\
Q^I_{~JK}(N)
   & \equiv & \biggl[
  -{V^I_{~JK} \over V} + {V^I_{~J}V_K \over V^2}
 + {V^I_{~K}V_J \over V^2} 
\cr &&\quad  
+ {V^I V_{JK} \over V^2} 
   - 2{V^I V_J V_K \over V^3}
    \biggr]_{\phi=\phizero(N)}. 
\end{eqnarray*}
Let us consider the conditions under which we can use Eqs.~(\ref{deltaphi1}) and (\ref{deltaphi2}). 
Differentiating Eq.~(\ref{deltaphi1}) with respect to $N$, we have
\begin{eqnarray}
{d \over dN} \left({d \over dN}\delta \phione^I\right) &=& {d \over dN}\left(\delta \phione^J\right) {P^{I}}_J
+\delta \phione^J{d \over dN}{P^I}_{J} \nonumber\\
&=& \delta \phione^J {\left(P^2\right)^I}_{J} - \delta \phione^J {Q^I}_{JK}{V^K \over V}~.
\label{difdeltaphi1}
\end{eqnarray}

If we consider the minus of Eq.~(\ref{difdeltaphi1}) as the corrections to the r.h.s. of Eq.~(\ref{deltaphi1}),
 we naively give an estimate for the correction to $\delta \phione^I$
\begin{eqnarray}
\Delta\left(\delta \phione^I\right) &\simeq& \int dN \delta \phione^J {P^I}_J + \int dN \delta \phione^J {\left(P^2\right)^I}_J 
\nonumber\\
&&\qquad\qquad
- \int \delta \phione^J {Q^I}_{JK}{V^K \over V}~.
\end{eqnarray} 
From this expression, in order that Eq.~(\ref{deltaphi1}) is a good approximation,
the conditions
\begin{eqnarray}
\left|\int dN {\left(P^2\right)^I}_{J} \right| \ll 1~,
\label{P2}
\end{eqnarray}
and
\begin{eqnarray}
\left|\int dN{Q^I}_{JK}\left({V^K \over V}\right)\right| \ll 1~,
\label{Q}
\end{eqnarray}
must be satisfied.
If we define a small parameter $\xi$ by
\begin{eqnarray*}
\left|{V^I \over V}\right| \equiv O(\xi)~,
\end{eqnarray*}
 we can estimate the duration of inflation measured in $N$ as  
\begin{eqnarray*}
\int dN \simeq \int dV {dV \over dN}^{-1} \sim O(\xi^{-2})~.
\end{eqnarray*}
Then, roughly speaking, the conditions,~(\ref{P2}) and (\ref{Q}), reduce to
\begin{eqnarray}
\left|{P^I}_J\right| \ll O(\xi)~,\quad
\left|{Q^I}_{JK}\right| \ll O(\xi)~.
\end{eqnarray}
Differentiating Eq.~(\ref{deltaphi2}) with respect to $N$, we also have
\begin{eqnarray*}
{d \over dN} \left({d \over dN}\delta \phitwo^I\right) &=& {d \over dN}\left(\delta \phitwo^J\right) {P^{I}}_J
+\delta \phitwo^J{d \over dN}{P^I}_{J} \nonumber\\
&& \quad + 2{d \over dN}\left(\delta \phione^J\right) \delta \phione^K {Q^I}_{JK} \nonumber\\
&& \quad + \delta \phione^J \delta \phione^K {d \over d\phi^L}\left({Q^I}_{JK}\right)\left(-{V^L \over V}\right)~.
\end{eqnarray*}
In the same way, the condition in which we can neglect the contribution of the second derivative, becomes 
\begin{eqnarray}
\left|{d \over d\phi^L}\left({Q^I}_{JK}\right)\right|\xi \ll \left|{Q^{I}}_{JK}\right|~.
\label{forth}
\end{eqnarray}

Thus, in order to use the approximate equations, Eqs.~(\ref{deltaphi1}) and (\ref{deltaphi2}), the required conditions are 
\begin{eqnarray}
&&\left|{V^I \over V} \right|= O(\xi)~,\label{first}\\
&&\left| {V_{IJ} \over V}\right| \sim \left|{V_{IJK} \over V} \right| \ll O(\xi)~,
\label{secondthird}
\end{eqnarray}
and Eq.~(\ref{forth})
for a small parameter $\xi$,
and
we redefine RSRC by these required conditions.

These equations are $D$-component coupled differential equations. 
In general, we cannot solve Eqs.~(\ref{deltaphi1}) and 
(\ref{deltaphi2}) analytically.
We give a formal solution of Eq.~(\ref{deltaphi1}) as 
\begin{equation}
\delta\phione^I(N) = \Lambda^I_{~J}(N,N_*)\delta\phi^J_*~,
\label{sol1}
\end{equation}
and 
\begin{equation}
\Lambda^I_{~J}(N,N') = 
\biggl[T\exp\biggl(\int^{N}_{N'}P(N'')dN''\biggr)\biggr]\!{{}^{}}^I_{~J}~,
\label{defLambda}
\end{equation}
where $T$ means the time-ordered product.
We also give the formal solution of Eq.~(\ref{deltaphi2}) as 
\begin{eqnarray}
\delta\phitwo^I(N) &\eq &
\Lambda^I_{~J}(N,N_*) \int^N_{N_*} dN'
 \left[\Lambda(N',N_*)^{-1}\right]^J_{~K}
\cr && \times
Q^K_{~LM}(N')
\delta\phione^L(N') \delta\phione^M(N')~.
\label{sol2}
\end{eqnarray}

Substituting these solutions to Eq.~(\ref{deltaNfend}), we obtain
\begin{eqnarray*}
\hspace{-5mm}\zeta(N_{\rm c}) 
&\eq & N_I^{\rm f} 
\biggl(\Lambda^I_{~J}(N_{\rm f},N_*)\delta\phi^J_* 
\cr &&
\qquad   +{1\over 2}\int_{N_*}^{N_{\rm f}} dN'\, 
   \Lambda^I_{~J}(N_{\rm f},N') Q^J_{~KL}(N')
\cr && \qquad\quad \times
  \Lambda^K_{~M}(N', N_*)\Lambda^L_{~N}(N', N_*)
     \delta\phi^M_*\delta\phi^N_*\biggr)\cr
  && +{1\over 2} N^{\rm f}_{IJ} 
   \Lambda^I_{~K}(N_{\rm f},N_*)\Lambda^J_{~L}(N_{\rm f},N_*)
     \delta\phi^K_* \delta\phi^L_*.  
\end{eqnarray*}
Comparing this expression with Eq.~(\ref{deltaNsecond}), 
we find that $N_I^*$ is expressed as 
\begin{eqnarray*}
N_I^* = N_J^{\rm f}\Lambda^J_{~I}(N_{\rm f},N_*). 
\end{eqnarray*}
Since the above relation should hold for arbitrary $N_*$, 
we also have 
\begin{eqnarray*}
 N_I(N)=N_J^{\rm f}\Lambda^J_{~I}(N_{\rm f},N).
\end{eqnarray*}
With the aid of this relation, $N_{IJ}^*$ is expressed 
as 
\begin{eqnarray*} 
N_{IJ}^* 
&\eq & 
  N^{\rm f}_{KL} 
   \Lambda^K_{~I}(N_{\rm f},N_*)\Lambda^L_{~J}(N_{\rm f},N_*)
\cr && 
+\int_{N_*}^{N_{\rm f}} dN'\, 
   N_K(N') Q^K_{~LM}(N')
\cr && \qquad\qquad \times
   \Lambda^L_{~I}(N', N_*)\Lambda^M_{~J}(N', N_*).
\end{eqnarray*}
Substituting the above relations into Eq.~(\ref{fNLdeltaN}), 
and using $\langle \phi_*^I \phi_*^J\rangle \propto \delta^{IJ}$, 
we finally obtain a very concise formula for the non-linear parameter
\begin{eqnarray}
 -{6 \over 5} {f_{NL}} &\eq &
   (N^*_I N_*^I)^{-2}\biggl(
  N^{\rm f}_{JK} 
   \Theta^J(N_{\rm f})\Theta^K(N_{\rm f})
\cr & +\!\!\!\! &
  \int_{N_*}^{N_{\rm f}}\!\! dN'\, 
   N_J(N') Q^J_{~KL}(N')
   \Theta^K(N')\Theta^L(N')
    \biggr),
\cr && 
\label{main}
\end{eqnarray}
where we have introduced a new vector 
\begin{equation}
\Theta^I(N) \equiv \Lambda^I_{~J}(N,N_*) N^J_*. 
\nonumber
\end{equation}

Eq.~(\ref{main}) is the main result of this paper.
If we directly evaluate the formula (\ref{fNLdeltaN})
for the non-linear parameter $f_{\rm NL}$, 
we need to calculate $\phitwo^I(N_{\rm f})$ 
as functions of $\phi_*^J$. 
Namely, we need to compute 
the coefficients $\phitwo^I_{~JK}(N_{\rm f})$ 
defined by $\phitwo^I(N)=\phitwo^I_{~JK}(N)\phi_*^J\phi_*^K$. 
However, our final expression (\ref{main}) does not 
request to compute the evolution of 
such a quantity that has three indices. 
Instead, we only have to deal with  
vector-like quantities $N_I(N)$ and $\Theta^I(N)$, which 
are obtained by solving 
\begin{eqnarray}
{d \over dN} N_I(N) &\eq & - P^J_{~I}(N) N_J(N)~,\\
{d \over dN} \Theta^I(N) &\eq & P^I_{~J}(N) \Theta^J(N)~. 
\label{ThetaEq}
\end{eqnarray}
The boundary condition for $N_I(N)$ is given at $N=N_{\rm f}$ 
and that for $\Theta^I(N)$ is given by 
\begin{eqnarray}
\Theta^J(N_*) = N^J (N_*)~. 
\label{Thetaini}
\end{eqnarray}
When a specific model is concerned, one can numerically evaluate 
$f_{NL}$ by using the above formula rather easily. 
We first numerically integrate $N_I(N)$ backwards in time 
until the initial time $N_*$, then the initial condition 
for $\Theta^I(N)$ is given by (\ref{Thetaini}). Solving 
Eq.~(\ref{ThetaEq}), we obtain $\Theta^I(N)$. Finally, substituting 
these results into the formula (\ref{main}) and integrating 
over $N'$, one obtain $f_{NL}$.

\subsection{non-linearity generated until $N=N_{\rm f}$}
\label{nconstant}

Here we evaluate $\zeta(N_{\rm f})$, 
the curvature perturbation on a uniform density hypersurface 
evaluated at $N=N_{\rm f}$. 
To do this, we use the following fact. 
It has been shown that perturbation of the background trajectories 
$\delta\phi^I(N_{\rm f})$ 
can be interpreted as the perturbation in a particular gauge, which 
we call $N$-constant gauge~\cite{Sasaki:1998ug}. 
Furthermore, it has been also shown that, under the assumption that 
one can neglect purely decaying mode contribution, 
$N$-constant gauge is equivalent to the flat slicing. 
Therefore 
$\delta\phi^I(N_{\rm f})$ can be recognized as 
the field perturbation on the flat slicing at $N=N_{\rm f}$. 
In the slow-roll regime, the uniform energy density hypersurface 
is approximately the same as the $V={\rm constant}$ hypersurface, 
since $\rho = V + {\cal O}(\epsilon)$.
Then, $\zeta(N_{\rm f})$ is evaluated by the time shift $\delta N$ 
necessary to transform to the $V={\rm constant}$ hypersurface 
from the flat slicing. 
This leads to the relation 
\begin{equation}
V\left(\,\phi(N_{\rm f}+\delta N)\right) = 
V\left(\,\phizero(N_{\rm f})\right)~.
\label{constantpotential}
\end{equation}
From this equation, we can obtain the relation between 
$\delta N=\zeta(N_{\rm f})$ and $\delta \phi^I_{\rm f}$.
Up to second order, Eq.~(\ref{constantpotential}) can be 
expanded as
\begin{eqnarray}
&& \hspace{-5mm}
  \left({V_{IJ}V^I V^J \over V^2}-{P^I_{~J}V_I V^J \over V}\right)
  \delta N^2 
\cr &&\quad
- 2\left({V^I V_I \over V} +V_{IJ}{V^I \over V}
  \delta\phione^J_{\rm f} 
   -V_I {d \over dN}\delta\phione^I_{\rm f}\right) 
  \delta N 
\cr &&\qquad
\left.+2V_I\delta\phione^{I}_{\rm f} + V_I\delta\phitwo^I_{\rm f} 
  + V_{IJ}\delta \phione^I_{\rm f} \delta\phione^J_{\rm f}
   \right\vert_{\phi=\phizero_{\rm f}}\!\!\!=0,\cr
&& 
\label{sod}
\end{eqnarray}
where we have used the equation of motion for $\phi^I(N)$, 
Eq.~(\ref{eom}), and its time derivative
\begin{equation}
{d^2 \over dN^2}\phi^I 
= - P^I_{~J}{V^J \over V}~.
\nonumber
\end{equation}
Solving Eq.~(\ref{sod}) for $\delta N$ up to second order 
in $\delta \phi$, we have
\begin{eqnarray}
&&\hspace{-4mm}\zeta(N_{\rm f}) \approx \delta N \cr
&& =
  {V \over V^I V_I} 
 \Biggl[V_J \delta\phione^J_{\rm f} 
+ {1\over 2}V_J\delta\phitwo^J_{\rm f}
+ {1 \over 2}U_{MN}
  \delta\phione^M_{\rm f} \delta\phione^N_{\rm f} 
\Biggr]_{\phi=\phizero_{\rm f}}, \cr\cr &&
\label{zetaNf}
\end{eqnarray}
with 
\begin{eqnarray*}
U_{MN} &\!\!\! \equiv \!\!\!& V_{MN} + 
  2{V_{KL}V^K V^L V_M V_N\over \left(V_J V^J \right)^2} 
\cr &&
\qquad\qquad + {V_M V_N \over V} -4 {V_{K(M} V^K V_N) \over V^J V_J},
\end{eqnarray*}
where we have used Eq.~(\ref{deltaphi1}).

From this expression, we find that one can apply the formula 
obtained in the preceding subsection with the identification 
\begin{eqnarray*}
 && 
N_I^{\rm f} = \left.\left({V \over V^J V_J}\right)V_I
    \right\vert_{\phi=\phizero_{\rm f}}~,
\cr &&
N_{IJ}^{\rm f} = \left.{1\over 2}
\left({V \over V^K V_K}\right)U_{IJ}
   \right\vert_{\phi=\phizero_{\rm f}}~. 
\end{eqnarray*}
In the present case 
the analytic formula for the non-linear parameter, $f_{NL}$,  
can be written down more explicitly as
\begin{eqnarray}
 -{6 \over 5} {f_{NL}} 
&\eq &2 \biggl[\epsilon
  + {\eta_{IJ}\over 2 V^K V_K}\bigl(
   2 V^I V^J
\cr &&\qquad\qquad\qquad
   -4 V^I \tilde\Theta^J 
   +\tilde\Theta^I \tilde\Theta^J 
   \bigr)\biggr]_{\phi=\phizero_{\rm f}}\cr
&&
 + (N^I_* N_I^*)^{-2}\int^{N_{\rm f}}_{N_*}
    dN\, N_I(N) Q^I_{~JK}(N)
\cr &&\qquad\qquad\qquad\qquad\times 
     \Theta^J(N)\Theta^K(N)~,
\label{main2}
\end{eqnarray}
with
\begin{equation}
\tilde\Theta^I=\left(N^K_* N_K^*\right)^{-1}V \Theta^I. 
\nonumber
\end{equation}
Here we have used the relation $\tilde \Theta^I V_I 
= V^I V_I$, which holds at $\phi^I=\phizero^I_{\rm f}$. 
In the single field case this result, Eq.~(\ref{main2}) corresponds to 
the well-known simple formula given in Ref~\cite{Lyth:2005du}
(see Appendix \ref{app}).
Under the condition that $N$-constant hypersurface is identical to 
$V=$constant hypersurface, it can be shown that $\zeta(N_{\rm f})$ 
becomes independent of $N_{\rm f}$, and so is 
$f_{NL}$. (see Appendix \ref{dedn}).  

We discuss here the rough order estimate of the above expression.
Here we assume Eqs.~(\ref{forth}),~(\ref{first}) and (\ref{secondthird}) for the order of magnitude 
of the derivative of the potential.
 Namely, $V_{IJ}/V \ll O(\xi)$ 
and $V_{IJK}/V \ll O(\xi)$. 
The duration of inflation measured in $N$ will be estimated by 
$V({dV/dN})^{-1}=O(\xi^{-2})$. 
Further, we assume that 
all components of $\Lambda^I_{~J}$ do not become much larger than unity. 
Then, $N_I$ and $\Theta_I$ are estimated as 
$O(\xi^{-1})$ since they are roughly the same order as $V/V_I$.
$\tilde\Theta_I$ is of $O(V\xi)$ and thence of $O(V^I)$. 
This rough estimate of the order of magnitude indicates that 
$f_{NL}$ is smaller than $O(1)$ within the range of validity of our  present approximation.  
The large non-Gaussianity ($f_{NL} \ge 1$) is not likely to be generated 
even in the case of multi-scalar inflation with non-separable potential in the standard slow-roll approximation, where
$V_{IJ}/V = O(\epsilon)$ and $V_{IJK}/V = O(\epsilon^{3/2})$,
$f_{NL}$ is definitely suppressed by the slow-roll parameter, $\epsilon$.
Nevertheless, a little loophole exists in the above estimate. 
We assumed that ${\Lambda^I}_{^J}$ is always of $O(1)$. 
However, since the exponent in Eq.~(\ref{defLambda}) can be $O(1)$,  
${\Lambda^I}_{^J}$ is not guaranteed to stay of $O(1)$.

\section{discussion \& conclusion}
\label{conclusion}
We have studied the primordial non-Gaussianity in $D$-component 
scalar field inflation models for 
arbitrary potential under the slow-roll approximation by making use
of the $\delta N$ formalism. 
We obtained a concise analytic formula for the non-linear 
parameter, $f_{NL}$, written in terms of the potential of 
the scalar field.
The obtained formula~(\ref{main}) has non-local terms, 
the terms written by an integral over $N'$ and the 
terms containing $\Theta^I$, which implicitly 
contains partial information of 
$\Lambda\sim T\exp\left( \int dN' P\right)$.  
Here $\Lambda$ and $P$ are a $D\times D$ matrix and 
$T$ represents the time-ordered product. 
Explicit form of $P^I_{~J}$ is given in (\ref{deltaphi1}). 
It is remarkable simplification that our final expression (\ref{main}) is written 
by 
$D$ vector quantities, $N_I$ and $\Theta^I$. 
Our formula is valid when $\eta_{IJ}\equiv V_{IJ}/V \ll O(\xi)$ and 
$V_{IJK}/V \ll O(\xi)$, where we defined $\xi$ by $V_I/V =O(\xi)$.  
In this case, we find that 
$f_{NL}$ is smaller than $O(1)$,
under the assumption that 
$N_I$ and $\Theta^I$ stays of $O(\xi^{-1})$,
even in the case of multi-scalar inflation with non-separable potential.
We also find that $f_{NL}$ is suppressed by the slow-roll parameter, $\epsilon$,
in the standard slow-roll inflation, where $\eta_{IJ} = O(\epsilon)$ and $V_{IJK}/V = O(\epsilon^{3/2})$. 
Under this assumption, 
primordial non-Gaussianity does not become large enough to be detectable
by future satellite missions for the cosmic microwave background 
($f_{NL} \ge 1$), such as PLANCK.

However, it is not clear if this assumption is guaranteed 
in general. In the standard slow-roll inflation, the 
exponent in the expression of $\Lambda$ also becomes $O(1)$. 
Hence, $\Lambda$ itself can be much larger than unity. 
As a future work, we will investigate various possibilities 
of generating large non-Gaussianity in multi-scalar inflation by 
constructing explicit models. 

Another possibility to generate detectable non-Gaussianity 
is to relax the conditions, $\eta_{IJ} \ll O(\xi)$ and $V_{IJK}/V \ll O(\xi)$. 
Observations currently constrain the magnitudes of 
the first and second derivatives of the potential in the direction along the 
background trajectory,
but the second derivatives of the potential in the direction
orthogonal to the background trajectory or
the third derivatives of the potential might be larger. 

In this paper we imposed the conditions that 
the derivatives of potential in all directions in field space 
are sufficiently small. 
We can relax these conditions still keeping
all the observational constraints satisfied.
We will also analyze such possibilities in a future work
by extending our formalism to non slow-roll cases.

\begin{acknowledgments}
TT is supported 
by Monbukagakusho Grant-in-Aid
for Scientific Research Nos.~17340075 and~19540285. 
This work is also supported in part by the 21st Century COE 
``Center for Diversity and Universality in Physics'' at Kyoto
 university, from the Ministry of Education,
Culture, Sports, Science and Technology of Japan. 

\end{acknowledgments}

\appendix

\section{Single field slow-roll case}
\label{app}
In the single field slow-roll case, the power spectrum is given by 
\begin{eqnarray*}
{\cal P}_{\zeta} = N_{\phi}^2{\cal P}_{*} = \left({V\Lambda \over V_{\phi}}\right)^2 \left({H \over 2\pi}\right)^2~.
\end{eqnarray*}
In this case, using the slow-roll parameters we can write $\Lambda$ as 
\begin{eqnarray*}
\Lambda = \exp\left[\int^{N}_{N_*}
  dN'\left(2\epsilon-\eta\right)\right]~.
\end{eqnarray*}
Moreover, we have
\begin{eqnarray*}
{d \over dN} V(N) &\eq & -2\epsilon V(N)~, \\
{d \over dN} V_{\phi}(N) &\eq & -\eta V_{\phi}(N)~.
\end{eqnarray*}
Using these equations, we obtain 
$V(N)=V^*$ $\exp(-2 \int^N_{N_*} \epsilon\, dN' 
)$ and $V_\phi(N)=V_\phi^*\exp( - \int^N_{N_*} \eta\, dN' 
)$. Then 
the power spectrum evaluated at horizon crossing time becomes  
\begin{eqnarray*}
{\cal P}_{\zeta} 
= \left({V_* \over V_{\phi *}}\right) \left({H_* \over 2\pi}\right)^2
= \left({H^2_* \over 2\pi \dot{\phi}_*}\right)^2~.
\end{eqnarray*}
This is consistent with a standard formula. 

In the single field slow-roll case, the non-linear parameter, ${f_{NL}}$
is simply given by 
\begin{eqnarray*}
-{6 \over 5}{f_{NL}} &\eq & {N_{\phi\phi} \over N_{\phi}^2}~, 
\end{eqnarray*}
where $N_{\phi} = \partial N / \partial \phi$. 
Since the 
e-folding number is given by 
\begin{eqnarray*}
N = - \int^{\phi_e}_{\phi_*}{V \over V_{\phi}}d\phi~,
\end{eqnarray*}
we obtain
\begin{eqnarray*}
N_{\phi} = {V \over V_{\phi}}~,~~N_{\phi\phi} = 1-{V V_{\phi\phi} \over V_{\phi}^2}~.
\end{eqnarray*}
Thus we have a simple formula as~\cite{Lyth:2005du} 
\begin{equation}
-{6 \over 5}{f_{NL}}={V_{\phi}^2 \over V^2} - {V_{\phi\phi} \over V} = 2\epsilon - \eta~.
\label{ssrfNL}
\end{equation}

Let us reproduce the same result by 
using our formula.  In the single field case 
our formula~(\ref{main2}) reduces to 
\begin{eqnarray*}
-{6 \over 5}{f_{NL}}&\eq & {V_{\phi}^2(N) \over V^2(N)} - {V_{\phi\phi}(N) \over V(N)}
 + {V_{\phi} \over V \Lambda} \int^{N}_0 dN' Q(N')\Lambda(N') \nonumber\\
&\eq & 2\epsilon(N) -\eta(N) + \int^{N}_{0}dN'\left[{d\eta \over dN'} -2{d\epsilon \over dN'}\right] \nonumber\\
&\eq & 2\epsilon_* - \eta_*~.
\end{eqnarray*}
This agrees with the formula given by Eq.~(\ref{ssrfNL}).

\section{Constancy of $\delta N$}
\label{dedn}

In the subsection \ref{nconstant}, we have evaluated $\zeta(N_{\rm f})$, 
the curvature perturbation on a
uniform density hypersurface evaluated at $N=N_{\rm {\rm f}}$. 
If $N(\phi,N_{\rm c})=$constant surface is identical to $V-$constant 
surface at $N=N_{\rm f}$, we have $\zeta(N_{\rm f})=\zeta(N_{\rm c})$. 
If it is always the case that $N-$constant surfaces in the configuration 
space are identical to $V-$constant ones
around $\phi^I=\phizero^I(N_{\rm f})$, we have 
$\zeta(N)=\zeta(N_{\rm c})$ for any $N$ close to $N_{\rm f}$. 
Hence, $\zeta(N)$ becomes independent of $N$. 

This can be directly verified as follows. 
In this case, the derivative of 
$\zeta(N_{\rm f})$ given in (\ref{zetaNf}) 
with respect $N_{\rm f}$,
\begin{eqnarray}
\dot\zeta(N) &\eq & \left(\dot{N_{I}} + N_J {P^J}_{I}\right) 
\left({\delta \phi^I}^{(1)} + {1 \over 2}{\delta \phi^J}^{(2)}\right) \nonumber\\
&&
+ {1 \over 2}\biggl[\dot{N_{IJ}} + {Q^K}_{IJ}N_{K} + N_{IK}{P^{K}}_{J} 
\cr && \qquad \qquad +
N_{JK}{P^{K}}_{I} \biggr]
\delta {\phi^I_{\rm f}}^{(1)} \delta {\phi^J_{\rm f}}^{(1)}~, 
\label{derideltan}
\end{eqnarray}
will be constant independent of $N$,  
where we have replaced $N_{\rm f}$ with $N$. 
The change rate of $V$ 
must be constant on $N-$constant surface in the present case. 
This condition can be written as 
\begin{equation}
{d \over d\phi^J}\left({{V'}^2 \over V}\right)\left({\delta^J}_K - {V^JV_K \over
 {V'}^2}\right) = 0, 
\nonumber
\end{equation}
which is further rewritten as
\begin{equation}
V_{IJ}V^IV^JV_K - {V'}^2V^IV_{IK} = 0.
\label{condition2}
\end{equation}
Using the identity $N_I\dot\phi^I=-1$ and 
$N_I\propto V_I$ which immediately follows from the fact that 
$N-$constant surfaces and $V-$constant surfaces are identical, 
we find that 
\begin{eqnarray*}
N_{I} = \left({V \over {V'}^2}\right)V_I~.  
\end{eqnarray*}
Then, we have  
\begin{eqnarray*}
\dot{N_{I}} + N_J {P^J}_{I} 
 = {2 \over {V'}^4}\left(V_{JK}V^JV^KV_I -
	 V_{IK}V^K\right) =0 ~,
\end{eqnarray*}
where we used the condition~(\ref{condition2}). 
Differentiation of the above equality gives 
\begin{eqnarray}
0&\eq&{d \over d\phi^K}\left(\dot{N_{I}} + N_J {P^J}_{I}\right)\cr
&\eq&
\dot{N_{IJK}}{d\phi^J \over dN} +{N_{IJ}}{d \over d\phi^K}\dot{\phi}^J + N_{JK}{P^J}_I
+N_{J}{Q^J}_{IK}
 \nonumber\\
&\eq&
\left(\dot{N_{IK}} + N_{IJ}{P^J}_{K} + N_{KJ}{P^{J}}_I  + {Q^J}_{IK}N_{J} \right)~.
\end{eqnarray}
From these relations, we can explicitly see that $\zeta(N)$ is
independent of $N$ when $N-$constant surfaces agree with $V-$constant 
surfaces. When this condition is satisfied, 
Eq.~(\ref{main2}) is slightly simplified. 
By using Eq.~(\ref{condition2}), 
the first two terms in round brackets can be unified 
into one term.

\end{document}